\documentclass[prm,twocolumn, amsmath,amssymb,
reprint,%
author-year,%
author-numerical,%
dvipdfmx
]{revtex4-1}

\usepackage{graphicx}
\usepackage{bm}
\usepackage{color}

\begin{document}


\title{Two-dimensional ionic crystals: The cases of IA-VII alkali halides and IA-IB CsAu}

\author{Shota Ono}
\email{shota\_o@gifu-u.ac.jp}
\affiliation{Department of Electrical, Electronic and Computer Engineering, Gifu University, Gifu 501-1193, Japan}

\begin{abstract}
The alkali halides, known as ionic crystals, have the NaCl-type or CsCl-type structure as the ground state. We study the structural, vibrational, and electronic properties of two-dimensional (2D) ionic crystals from first-principles. Two potential structures that are hexagonal and tetragonal are investigated as structural templates. Through phonon dispersion calculations, 8 and 16 out of 20 alkali halides in the hexagonal and tetragonal structures are dynamically stable, respectively. The electron energy gaps range from 6.8 eV for LiF to 3.9 eV for RbI and CsI in the tetragonal structure within the generalized gradient approximation. By considering the Madelung energy and the core-core repulsion, we propose a hard sphere model that accounts for the nearest-neighbor bond length and the cohesive energy of 2D alkali halides. The 2D CsAu in the tetragonal structure is also predicted to be stable as an ionic crystal including only metallic elements, showing a band gap of 2.6 eV that is higher than that of the 3D counterparts. 
\end{abstract}

\maketitle

\section{Introduction}
The solids are classified into metals and insulators, and the latter is further classified into covalent, ionic, and molecular crystals. The covalent bonding has played an important role in creating two-dimensional (2D) crystals, such as graphene, hexagonal boron-nitride, and transition metal dichalcogenides \cite{heine}. The honeycomb lattice with zero or finite thickness is a structural template in 2D covalent crystals, while the interlayer coupling is described by a van der Waals interaction. As the 3D metals have the fcc and hcp structures, the 2D metals can form close packed structure of hexagonal or buckled honeycomb lattices \cite{nevalaita,ono2020}, while most of them are left to be synthesized. The 2D ionic crystals have also attracted significant interest recently. Tikhomirova {\it et al}. predicted the stable structures of 2D NaCl by using density functional theory (DFT) and evolutionary approach, and synthesized 2D NaCl experimentally, where the stability of the 2D NaCl relies on the chemical interaction with the diamond substrate \cite{tikhomirova}. Shi {\it et al}. synthesized Na-rich phases of 2D Na-Cl crystals that are stabilized on graphene surface \cite{shi}. Zhao {\it et al}. reported the square lattice formation of 2D NaCl using molecular dynamics (MD) simulation \cite{zhao}. Chen {\it et al}. found 2D EuS as an ionic crystal having ferromagnetism within DFT \cite{cheng}.

Several calculations within DFT have been performed to predict stable structures and interesting properties of 2D compounds including group-IV \cite{sahin}, III-V \cite{sahin,hennig2013,hennig2014}, III-VI \cite{cahangirov}, II-VI \cite{II-V}, IV-VI \cite{IV-VI}, V-IV \cite{IV-V,ozdamar}, and V-IV-III-VI \cite{IV-V} elements. While the group-IV elements have perfect covalent bonds, any other compounds can have ionic bonds as well due to the electronegativity difference \cite{II-V}. In this sense, the 2D IA-VII compounds have strong ionic bonds. More recently, Kumar {\it et al}. investigated the structural and electronic properties of 2D alkali halides except for the Cs-based compounds, whereas only the planar honeycomb structure was assumed \cite{kumar}. 

The alkali halides as 3D ionic crystals can have the NaCl-type (B1) or CsCl-type (B2) structures in the ground state. Whereas the relative phase stability has been studied in detail using first-principles approach \cite{florez} and Landau free energy \cite{toledano}, such a preference can be simply determined by the interplay between the Madelung energy and the ratio of the ionic radii $R=r^>/r^<$, where $r^>$ and $r^<$ are the radii of the larger and smaller ions, respectively \cite{maradudin,AM}. The Madelung constant of the CsCl-type structure ($\alpha_{\rm B2} \simeq 1.7627$) is larger than that of the NaCl-type structure ($\alpha_{\rm B1} \simeq 1.7476$), indicating that the former is more stable if the nearest-neighbor interatomic distance is the same in both structures. On the other hand, when the value of $R$ is increased, the stable structure changes from the CsCl-type to NaCl-type structure across the value of $R=(\sqrt{3}+1)/2\simeq 1.37$. This is speculated from the 3D packing of hard (i.e., impenetrable) spheres: the nearest-neighbor interatomic distance is $\sqrt{3}r^>$ (between the atoms with the same charges) rather than $r^>+r^<$ (between the atoms with different charges) \cite{pettifor}. For the 2D ionic crystals, however, no simple models have been established for understanding the stable structures. 

\begin{table*}
\begin{center}
\caption{The 2D $AB$ identified to be dynamically stable as the formula of $A_1B_1$ or $A_2B_2$. These $AB$s exhibit semiconducting property except for III-V GaSb \cite{hennig2013}, IV-IV C \cite{balendhran}, IV-V CBi, GeBi, SnBi, PbN, PbSb \cite{ozdamar}, and bimetallic systems M-M \cite{ono_satomi,ono2021PRM}. For the IV-IV compounds, the cases of $A=B$ indicate graphene, silicene, germanene, and stanene for $A=$ C, Si, Ge, and Sn, respectively.}
{
\begin{tabular}{llr}\hline\hline
 group  \hspace{5mm} & compounds  & source  \\  \hline
 IA-VII & LiF, LiCl, LiBr, NaF, NaCl, KF, KCl, KBr, KI, RbF, RbCl, RbBr, RbI, CsF, CsCl, CsBr, CsI  &  this work \\
 IA-IB & CsAu &  this work \\
 II-V & BeO, MgO, CaO, CaS, SrS, SrSe, BaS, BaTe, ZnO, CdO, HgS, HgTe & \cite{II-V} \\
 III-V & $A =$ Al, Ga, In; $B =$ N, P, As, Sb  & \cite{sahin,hennig2013} \\ 
 III-VI & $A =$ B, Al, Ga, In; $B=$ O, S, Se, Te & \cite{cahangirov} \\
 IV-IV & $A, B =$ C, Si, Ge, Sn  & \cite{sahin,balendhran} \\
 IV-V & $A =$ C, Si, Ge, Sn, Pb; $B =$ N, P, As, Sb, Bi & \cite{IV-V,ozdamar} \\
 M-M & 15 Li$B$, 14 Cu$B$, 17 Ag$B$, 19 Au$B$, 7 Ir$B$, and a few Mg$B$, Ru$B$, Rh$B$, Os$B$ ($B=$ metallic elements) & \cite{ono_satomi,ono2021PRM} \\
\hline\hline
\end{tabular}
}
\label{table_2D}
\end{center}
\end{table*}

\begin{figure}
\center
\includegraphics[scale=0.35]{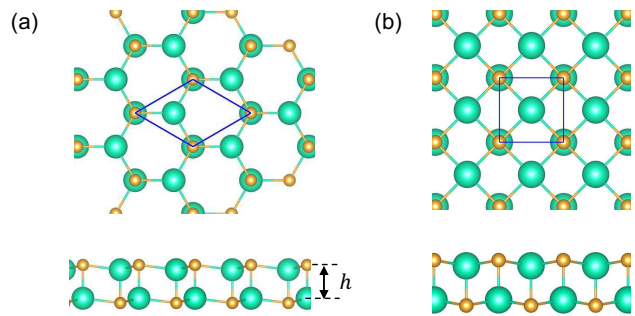}
\caption{Top and side views of 2D $AB$ in the (a) hexagonal and (b) tetragonal structures. Solid line (blue) indicates the unit cells that include four atoms. The thickness of the 2D $AB$ is indicated by $h$. } \label{fig_1} 
\end{figure}

In this paper, we explore the stable structures of 2D alkali halides $AB$ (IA-VII compounds). Two crystal structures with the hexagonal and tetragonal lattices are assumed, as shown in Fig.~\ref{fig_1}. These structures have been identified as the 2D CuI-type and SnSe-type structures, respectively, within a computational search based on a topology-scaling algorithm \cite{ashton}, and the latter structure has also been assumed in studying 2D ionic EuS crystals \cite{cheng}. Based on the Madelung constant calculations and the 2D sphere packing, we predict that the tetragonal structure is more stable than the hexagonal structure for $R \le \sqrt{2}+1$. By using DFT, we confirm that the 2D alkali halides $AB$ ($A=$ Li, Na, K, Rb, Cs; $B=$ F, Cl, Br, I) follow this trend, that is, the 2D alkali halides have the tetragonal phase except for the Li-based compounds. We demonstrate that 8 and 16 out of 20 $AB$ in the hexagonal and tetragonal structures, respectively, are dynamically stable by performing phonon dispersion calculations. The 2D alkali halides have an electron energy gap that ranges from 6.8 eV for LiF to 3.9 eV for RbI and CsI in the tetragonal structure, and exhibit a relatively flat band near the Fermi level. We also predict that the 2D tetragonal CsAu, an ionic crystal including only metallic elements, have an energy gap of 2.6 eV that is higher than that of the 3D counterpart by a factor of more than two.

In Table~\ref{table_2D}, we list the 2D $AB$s that have been identified to be dynamically stable within phonon calculations. We also note that the Computational 2D Materials Database (C2DB) \cite{C2DB2021}, created from high-throughput DFT calculations, includes 225 $AB$ compounds that are dynamically stable. The C2DB includes 8 alkali halides having the formula of $A_2B_2$ ($A=$ K, Rb, Cs; $B=$ Cl, Br, I) except for Cs$_2$I$_2$. The stable structure is the same as the 2D CuBr-type structure \cite{ashton}: the Cu and Br atoms form the square structure, where the Br atoms are displaced alternately above and below the plane of Cu atoms. The present work is expected to stimulate more investigations for understanding the stabilities of 2D ionic crystals. 

The paper is organized as follows: in Sec.~\ref{sec:results} we first discuss the stable structures based on a hard sphere model (Sec.~\ref{sec:hard_sphere}), next provide the structural, vibrational, and electronic properties of 2D alkali halides through first-principles calculations (Sec.~\ref{sec:dft_dfpt}), and compare the cohesive energies within DFT to those within a hard sphere model (Sec.~\ref{sec:dft_hard_sphere}). The physical properties of 2D CsAu are provided in Sec.~\ref{sec:csau}, and our conclusion is presented in Sec.~\ref{sec:conclusion}. The computational details are included in Appendix.

\section{Results and Discussion}
\label{sec:results}
\subsection{Hard sphere model}
\label{sec:hard_sphere}
Figure \ref{fig_1} shows the geometry of the hexagonal and tetragonal structures assumed in the present work. The four atoms ($A_2B_2$) are included in the unit cell; two of which ($A_1B_1$) form a monolayer in the honeycomb or square structures; and those monolayers are stacked along the surface normal to form the $AB$ bonds. Such an ionic bonding along the $z$ direction is expected to stabilize the 2D crystals against the out-of-plane vibrations. The basis vectors for the $A1$ and $B1$ atoms are $\bm{\tau}_{A1}=(0,0,-z_A)$ and $\bm{\tau}_{B1}=(0,0,z_B)$, respectively, while those for the $A2$ and $B2$ atoms are given by $\bm{\tau}_{A2}=(0,a/\sqrt{3},z_A)$ and $\bm{\tau}_{B2}=(0,a/\sqrt{3},-z_B)$ for the hexagonal structure and $\bm{\tau}_{A1}=(0,0,-z_A)$ $\bm{\tau}_{B1}=(0,0,z_B)$, $\bm{\tau}_{A2}=(a/2,a/2,z_A)$, and $\bm{\tau}_{B2}=(a/2,a/2,-z_B)$ for the tetragonal structure, respectively, with the lattice constant $a$. We define the thickness of the 2D materials as $h=z_A+z_B$. 

\begin{figure}
\center
\includegraphics[scale=0.45]{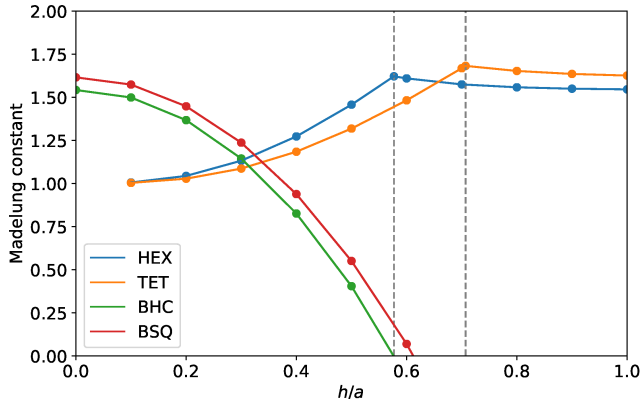}
\caption{The Madelung constant as a function of $h/a$ for the hexagonal (HEX), tetragonal (TET), buckled honeycomb (BHC), and buckled square (BSQ) structures. The optimal values of $h/a$ for the hexagonal ($1/\sqrt{3}\simeq 0.577$) and tetragonal ($1/\sqrt{2}\simeq 0.707$) structures are indicated by vertical lines. } \label{fig_2} 
\end{figure}

Figure \ref{fig_2} shows the $h$-dependence of the Madelung constant $\alpha_j$ for the 2D structures $j$ including the hexagonal (Fig.~\ref{fig_1}(a)) and tetragonal (Fig.~\ref{fig_1}(b)) structures, where the equality of $z_A=z_B$ is assumed. For comparison, two other structures were also assumed: the buckled honeycomb structure with $\bm{\tau}_{A}=(0,0,-z_A)$ and $\bm{\tau}_{B}=(0,a/\sqrt{3},z_B)$ and the buckled square structure with $\bm{\tau}_{A}=(0,0,-z_A)$ and $\bm{\tau}_{B}=(a/2,a/2,z_B)$ \cite{ono_satomi}. The values of $\alpha_j$ were calculated by using pymatgen code \cite{pymatgen}. The hexagonal and tetragonal structures have the maximum value of $\alpha_j \simeq 1.621$ at $h/a=1/\sqrt{3}$ and $\alpha_j\simeq 1.682$ at $h/a=1/\sqrt{2}$, respectively. The optimal values of $h/a$ are obtained when the interlayer distance is equal to the interatomic distance between atoms $A$ and $B$ in each monolayer. The buckled honeycomb and square structures have the maximum values of $\alpha_j \simeq 1.542$ and $1.615$ at $h/a=0$, respectively, so that the buckled structures with finite $h/a$ will be energetically unstable, which is in contrast to the structure-stability property of the 2D metals \cite{ono2020}. Since the Madelung energy is given by $E = - \alpha_j e^2 /d$ with the electron charge $e$ and the nearest-neighbor distance $d$, the 2D ionic crystals are predicted to have the tetragonal structure when the values of $d$ are the same for all $j$s. 

We next investigate a potential transformation from the tetragonal to hexagonal structures by considering the difference of the ionic radii between atoms $A$ and $B$ within a 2D sphere packing. For the tetragonal structure, the interatomic distance is equal to $d=r^>+r^<$. However, when $r^>=a/\sqrt{2}$, the smaller ions may not touch the larger ones, yielding $d=\sqrt{2}r^>$. Such a situation occurs at a critical ratio given by 
\begin{eqnarray}
 R_{\rm tet} = \sqrt{2} + 1\simeq 2.414. 
 \label{eq:C1}
\end{eqnarray}
In a similar manner, the critical ratio $R_{\rm hex}$ of the hexagonal structure is given by $R_{\rm hex} = \sqrt{3} (2+ \sqrt{3}) \simeq 6.464$. For the alkali halides $AB$, the values of $R$ are listed in Table \ref{table_R}. The Li atom has a small ionic radius of $0.6$ \AA, so that the $R$ is larger than $R_{\rm tet}$ except for LiF. Other alkali halides satisfies $R<R_{\rm tet}$. From the calculations of the $\alpha_j$ and $R$, we predict that the Na-, K-, Rb-, and Cs-based compounds (except for the Li-based compounds) prefer the tetragonal structure. 

\begin{figure}
\center
\includegraphics[scale=0.45]{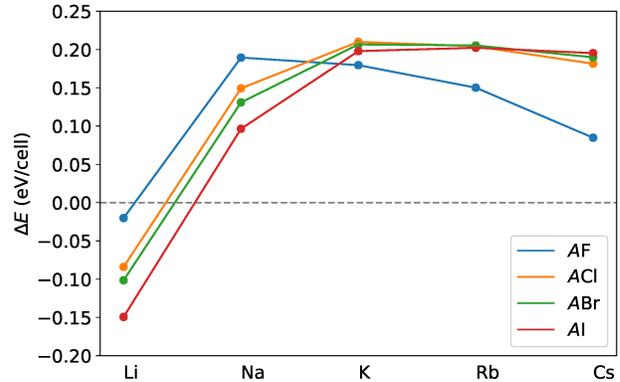}
\caption{The plot of $\Delta E$ for $AB$ ($A=$ Li, Na, K, Rb, Cs; $B=$ F, Cl, Br, I). Negative and positive values of $\Delta E$ indicate that the hexagonal and tetragonal structures are stable, respectively. } \label{fig_3} 
\end{figure}

\begin{table}
\begin{center}
\caption{The value of $R$ for the alkali halides $AB$. The figure of the parenthesis for the atoms $A$ and $B$ indicates the ionic radius (\AA). The data were extracted from Ref.~\cite{AM}. }
{
\begin{tabular}{lccccccccc}\hline\hline
   \hspace{5mm} & Li (0.60) & Na (0.95) & K (1.33) & Rb (1.48) & Cs (1.69)  \\  \hline
F \ (1.36) & 2.27 & 1.43 & 1.02 & 1.09 & 1.24  \\
Cl (1.81) & 3.02 & 1.91 & 1.36 & 1.22 & 1.07  \\
Br (1.95) &  3.25 & 2.05 & 1.47 & 1.32 & 1.15 \\
I \ \ (2.16) & 3.60 & 2.27 & 1.62 & 1.46 & 1.28  \\
\hline\hline
\end{tabular}
}
\label{table_R}
\end{center}
\end{table}

\begin{figure*}
\center
\includegraphics[scale=0.55]{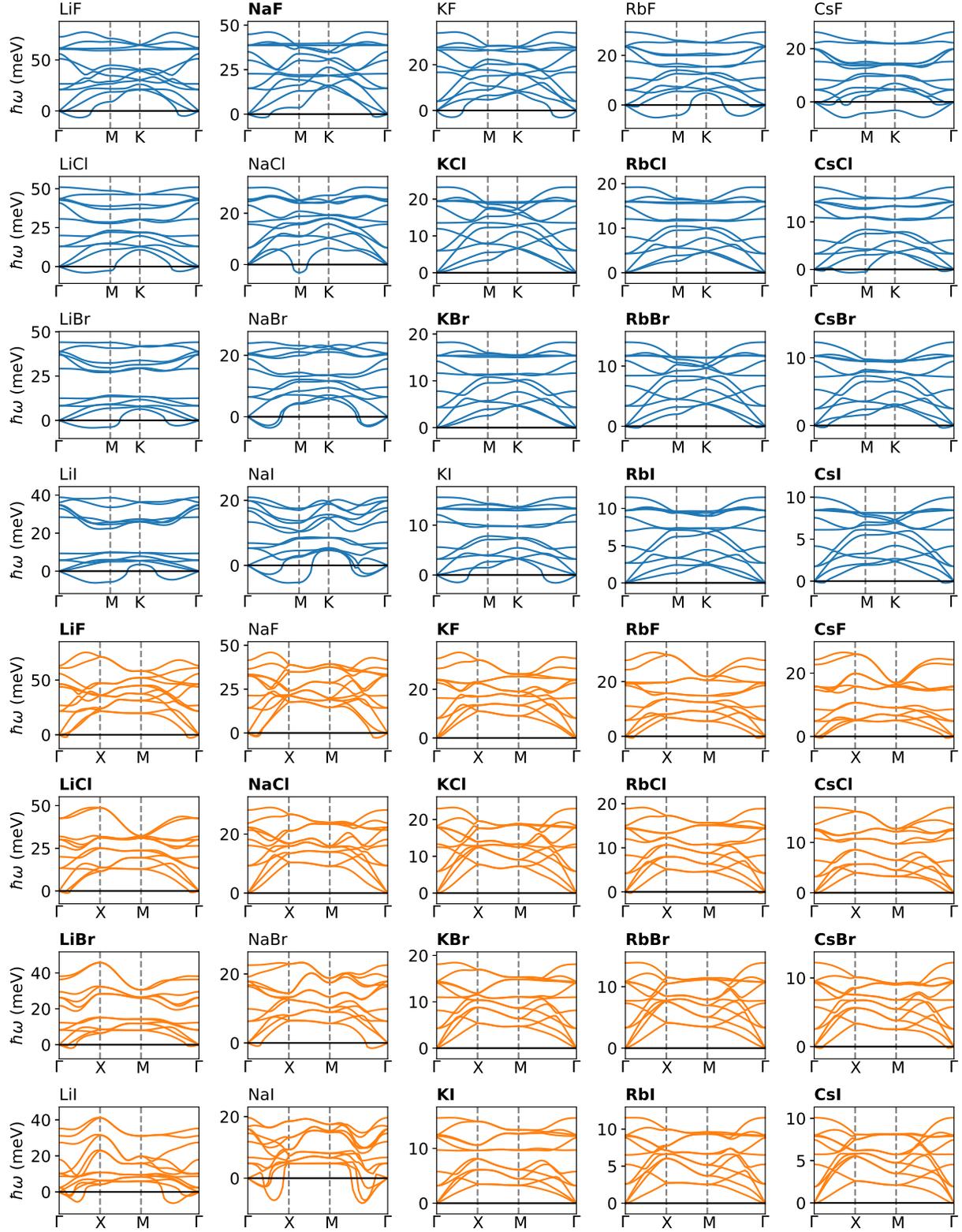}
\caption{The phonon dispersions of 2D alkali halides in the hexagonal (blue) and tetragonal (orange) structures. The imaginary frequencies are indicated by negative values. The dynamically stable $AB$ is shown in bold, where the ratio of the minimum and maximum frequencies is larger than $-0.05$.  } \label{fig_4} 
\end{figure*}

\begin{figure*}
\center
\includegraphics[scale=0.55]{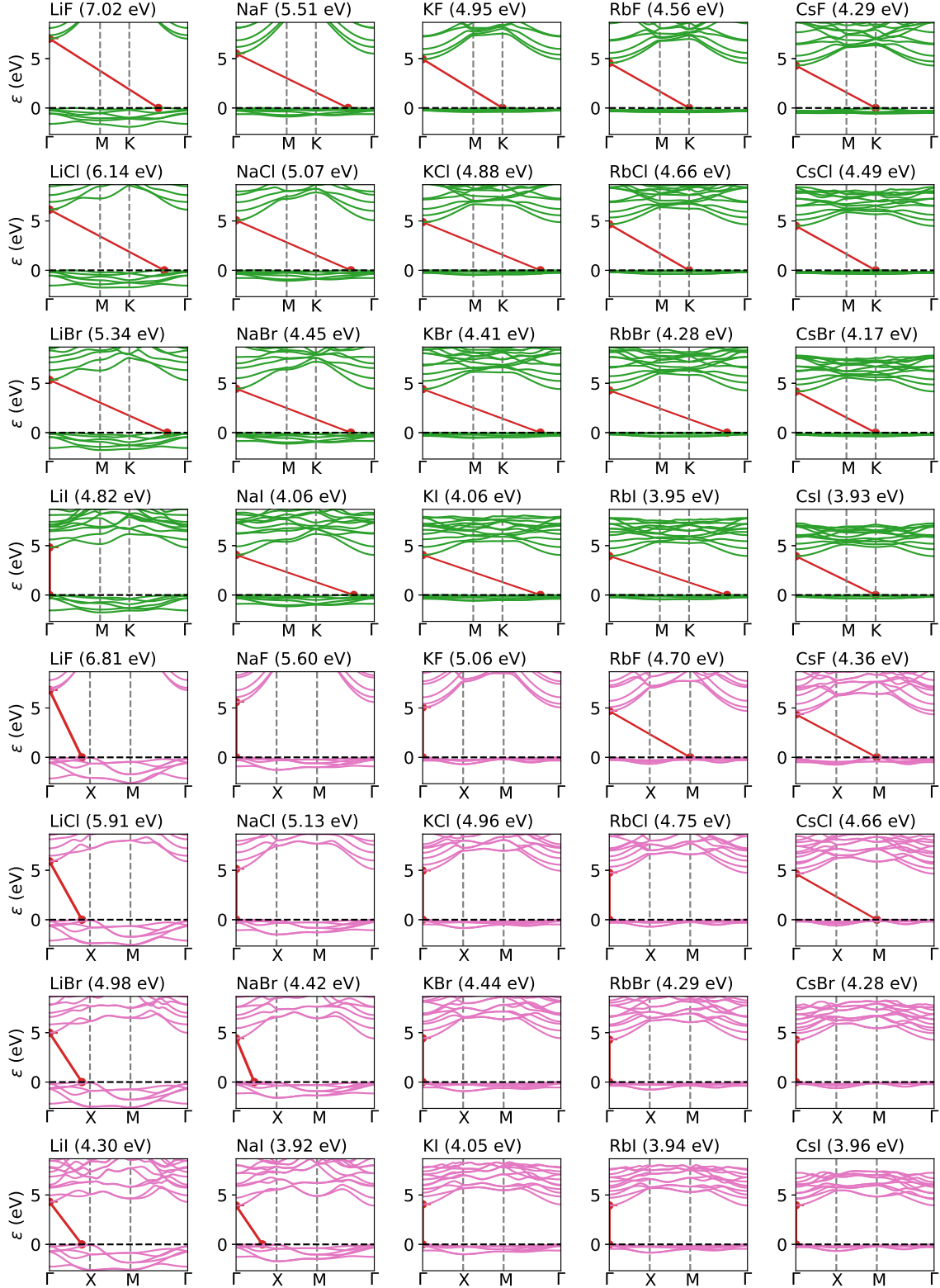}
\caption{The electron band structure of 2D alkali halides in the hexagonal (green) and tetragonal (pink) structures. The energy is measured from the Fermi level. The VBM and CBM are indicated by red circles connected by a straight line. The magnitude of the band gap is indicated in parenthesis. } \label{fig_5} 
\end{figure*}

\begin{table*}
\begin{center}
\caption{The lattice parameters of $a$, $z_A$, and $z_B$ (\AA) for the 2D $AB$ in the hexagonal and tetragonal structures. The elastic constants of $B_{\rm 2D}$, $c_{11}$, and $c_{12}$ (GPa nm) and the value of $\Delta$ for the tetragonal structure. }
{
\begin{tabular}{lcccccccccc}\hline\hline
 \hspace{3mm} & hexagonal  \hspace{3mm} & \hspace{3mm} &  \hspace{3mm} & tetragonal \hspace{3mm} &  \hspace{3mm} &  \hspace{3mm} &  \hspace{3mm} &  \hspace{3mm} &  \hspace{3mm} &   \\ 
$AB$ \hspace{3mm} & $a$ \hspace{3mm} & $z_{A}$ \hspace{3mm} & $z_{B}$ \hspace{3mm} & $a$ \hspace{3mm} & $z_A$ \hspace{3mm} & $z_B$ \hspace{3mm} & $B_{\rm 2D}$ \hspace{3mm} & $c_{11}$  \hspace{3mm} & $c_{12}$ \hspace{3mm} & ${\rm log}_{10}\Delta$  \\ \hline
LiF \hspace{4mm} & 3.19 \hspace{4mm} & 0.99 \hspace{4mm} & 1.08 \hspace{4mm} & 2.76 \hspace{4mm} & 1.00 \hspace{4mm} & 1.07 \hspace{4mm} & 45.4 \hspace{4mm} & 83.2 \hspace{4mm} & 7.1 \hspace{4mm} & 3.8 \\
LiCl \hspace{4mm} & 4.04 \hspace{4mm} & 1.07 \hspace{4mm} & 1.42 \hspace{4mm} & 3.52 \hspace{4mm} & 1.10 \hspace{4mm} & 1.38 \hspace{4mm} & 25.0 \hspace{4mm} & 43.7 \hspace{4mm} & 6.3 \hspace{4mm} & 3.3 \\
LiBr \hspace{4mm} & 4.30 \hspace{4mm} & 1.07 \hspace{4mm} & 1.55 \hspace{4mm} & 3.77 \hspace{4mm} & 1.12 \hspace{4mm} & 1.50 \hspace{4mm} & 21.2 \hspace{4mm} & 36.6 \hspace{4mm} & 5.6 \hspace{4mm} & 3.1 \\
LiI \hspace{4mm} & 4.65 \hspace{4mm} & 1.07 \hspace{4mm} & 1.76 \hspace{4mm} & 4.12 \hspace{4mm} & 1.15 \hspace{4mm} & 1.67 \hspace{4mm} & 17.2 \hspace{4mm} & 30.0 \hspace{4mm} & 4.2 \hspace{4mm} & 2.9 \\
NaF \hspace{4mm} & 3.81 \hspace{4mm} & 1.16 \hspace{4mm} & 1.19 \hspace{4mm} & 3.22 \hspace{4mm} & 1.18 \hspace{4mm} & 1.20 \hspace{4mm} & 37.6 \hspace{4mm} & 56.9 \hspace{4mm} & 18.5 \hspace{4mm} & 3.5 \\
NaCl \hspace{4mm} & 4.64 \hspace{4mm} & 1.31 \hspace{4mm} & 1.51 \hspace{4mm} & 3.93 \hspace{4mm} & 1.34 \hspace{4mm} & 1.50 \hspace{4mm} & 22.2 \hspace{4mm} & 32.4 \hspace{4mm} & 12.0 \hspace{4mm} & 3.0 \\
NaBr \hspace{4mm} & 4.91 \hspace{4mm} & 1.34 \hspace{4mm} & 1.63 \hspace{4mm} & 4.16 \hspace{4mm} & 1.37 \hspace{4mm} & 1.61 \hspace{4mm} & 18.9 \hspace{4mm} & 27.3 \hspace{4mm} & 10.5 \hspace{4mm} & 2.8 \\
NaI \hspace{4mm} & 5.29 \hspace{4mm} & 1.37 \hspace{4mm} & 1.81 \hspace{4mm} & 4.50 \hspace{4mm} & 1.42 \hspace{4mm} & 1.78 \hspace{4mm} & 15.2 \hspace{4mm} & 21.9 \hspace{4mm} & 8.5 \hspace{4mm} & 2.6 \\
KF \hspace{4mm} & 4.41 \hspace{4mm} & 1.36 \hspace{4mm} & 1.30 \hspace{4mm} & 3.73 \hspace{4mm} & 1.36 \hspace{4mm} & 1.32 \hspace{4mm} & 26.6 \hspace{4mm} & 37.4 \hspace{4mm} & 15.8 \hspace{4mm} & 3.1 \\
KCl \hspace{4mm} & 5.27 \hspace{4mm} & 1.53 \hspace{4mm} & 1.60 \hspace{4mm} & 4.42 \hspace{4mm} & 1.55 \hspace{4mm} & 1.62 \hspace{4mm} & 17.5 \hspace{4mm} & 23.6 \hspace{4mm} & 11.4 \hspace{4mm} & 2.6 \\
KBr \hspace{4mm} & 5.55 \hspace{4mm} & 1.58 \hspace{4mm} & 1.71 \hspace{4mm} & 4.65 \hspace{4mm} & 1.61 \hspace{4mm} & 1.72 \hspace{4mm} & 15.2 \hspace{4mm} & 20.2 \hspace{4mm} & 10.2 \hspace{4mm} & 2.5 \\
KI \hspace{4mm} & 5.96 \hspace{4mm} & 1.65 \hspace{4mm} & 1.88 \hspace{4mm} & 4.98 \hspace{4mm} & 1.69 \hspace{4mm} & 1.87 \hspace{4mm} & 12.7 \hspace{4mm} & 16.6 \hspace{4mm} & 8.8 \hspace{4mm} & 2.3 \\
RbF \hspace{4mm} & 4.66 \hspace{4mm} & 1.45 \hspace{4mm} & 1.34 \hspace{4mm} & 3.95 \hspace{4mm} & 1.46 \hspace{4mm} & 1.36 \hspace{4mm} & 23.6 \hspace{4mm} & 32.7 \hspace{4mm} & 14.6 \hspace{4mm} & 2.9 \\
RbCl \hspace{4mm} & 5.53 \hspace{4mm} & 1.63 \hspace{4mm} & 1.65 \hspace{4mm} & 4.65 \hspace{4mm} & 1.65 \hspace{4mm} & 1.67 \hspace{4mm} & 16.0 \hspace{4mm} & 21.1 \hspace{4mm} & 10.9 \hspace{4mm} & 2.5 \\
RbBr \hspace{4mm} & 5.81 \hspace{4mm} & 1.68 \hspace{4mm} & 1.76 \hspace{4mm} & 4.87 \hspace{4mm} & 1.71 \hspace{4mm} & 1.77 \hspace{4mm} & 14.0 \hspace{4mm} & 18.2 \hspace{4mm} & 9.8 \hspace{4mm} & 2.4 \\
RbI \hspace{4mm} & 6.22 \hspace{4mm} & 1.76 \hspace{4mm} & 1.92 \hspace{4mm} & 5.20 \hspace{4mm} & 1.79 \hspace{4mm} & 1.92 \hspace{4mm} & 11.8 \hspace{4mm} & 15.1 \hspace{4mm} & 8.6 \hspace{4mm} & 2.2 \\
CsF \hspace{4mm} & 4.92 \hspace{4mm} & 1.58 \hspace{4mm} & 1.34 \hspace{4mm} & 4.21 \hspace{4mm} & 1.57 \hspace{4mm} & 1.37 \hspace{4mm} & 19.0 \hspace{4mm} & 25.9 \hspace{4mm} & 12.0 \hspace{4mm} & 2.7 \\
CsCl \hspace{4mm} & 5.84 \hspace{4mm} & 1.74 \hspace{4mm} & 1.69 \hspace{4mm} & 4.91 \hspace{4mm} & 1.76 \hspace{4mm} & 1.71 \hspace{4mm} & 14.0 \hspace{4mm} & 18.1 \hspace{4mm} & 10.0 \hspace{4mm} & 2.4 \\
CsBr \hspace{4mm} & 6.12 \hspace{4mm} & 1.80 \hspace{4mm} & 1.80 \hspace{4mm} & 5.13 \hspace{4mm} & 1.82 \hspace{4mm} & 1.82 \hspace{4mm} & 12.6 \hspace{4mm} & 16.0 \hspace{4mm} & 9.1 \hspace{4mm} & 2.2 \\
CsI \hspace{4mm} & 6.53 \hspace{4mm} & 1.88 \hspace{4mm} & 1.96 \hspace{4mm} & 5.46 \hspace{4mm} & 1.91 \hspace{4mm} & 1.97 \hspace{4mm} & 10.8 \hspace{4mm} & 13.4 \hspace{4mm} & 8.2 \hspace{4mm} & 2.1 \\
\hline\hline
\end{tabular}
}
\label{table_dft}
\end{center}
\end{table*}

\begin{table}
\begin{center}
\caption{The values of $d$ in units of \AA, $E$ in units of eV/(ion pair), and $m$ for the 2D alkali halides $AB$ in the tetragonal structure. The subscript ``sp'' indicates the hard sphere model. }
{
\begin{tabular}{lccccccc}\hline\hline
$AB$ \hspace{3mm} & $d_{\rm DFT}$ \hspace{3mm} & $d_{\rm sp}$ \hspace{3mm} & $E_{\rm DFT}$ \hspace{3mm} & $E_{\rm sp}^{\rm coul}$ \hspace{3mm} & $m$ \hspace{3mm} & $E_{\rm sp}^{\rm tot}$\\ \hline
LiF \hspace{3mm} & 1.95 \hspace{3mm} & 1.96 \hspace{3mm} & 9.31 \hspace{3mm} & 12.36 \hspace{3mm} & 4.52 \hspace{3mm} & 9.63 \\
LiCl \hspace{3mm} & 2.51 \hspace{3mm} & 2.56 \hspace{3mm} & 7.10 \hspace{3mm} & 10.05 \hspace{3mm} & 5.07 \hspace{3mm} & 8.07 \\
LiBr \hspace{3mm} & 2.69 \hspace{3mm} & 2.76 \hspace{3mm} & 6.38 \hspace{3mm} & 9.50 \hspace{3mm} & 5.24 \hspace{3mm} & 7.68 \\
LiI \hspace{3mm} & 2.96 \hspace{3mm} & 3.05 \hspace{3mm} & 5.58 \hspace{3mm} & 8.77 \hspace{3mm} & 5.57 \hspace{3mm} & 7.20 \\
NaF \hspace{3mm} & 2.28 \hspace{3mm} & 2.31 \hspace{3mm} & 8.20 \hspace{3mm} & 10.48 \hspace{3mm} & 5.78 \hspace{3mm} & 8.67 \\
NaCl \hspace{3mm} & 2.78 \hspace{3mm} & 2.76 \hspace{3mm} & 6.43 \hspace{3mm} & 8.77 \hspace{3mm} & 5.81 \hspace{3mm} & 7.27 \\
NaBr \hspace{3mm} & 2.95 \hspace{3mm} & 2.90 \hspace{3mm} & 5.79 \hspace{3mm} & 8.35 \hspace{3mm} & 5.75 \hspace{3mm} & 6.90 \\
NaI \hspace{3mm} & 3.20 \hspace{3mm} & 3.11 \hspace{3mm} & 5.08 \hspace{3mm} & 7.79 \hspace{3mm} & 5.71 \hspace{3mm} & 6.42 \\
KF \hspace{3mm} & 2.64 \hspace{3mm} & 2.69 \hspace{3mm} & 7.93 \hspace{3mm} & 9.00 \hspace{3mm} & 6.34 \hspace{3mm} & 7.58 \\
KCl \hspace{3mm} & 3.13 \hspace{3mm} & 3.14 \hspace{3mm} & 6.44 \hspace{3mm} & 7.71 \hspace{3mm} & 6.59 \hspace{3mm} & 6.54 \\
KBr \hspace{3mm} & 3.29 \hspace{3mm} & 3.28 \hspace{3mm} & 5.86 \hspace{3mm} & 7.38 \hspace{3mm} & 6.53 \hspace{3mm} & 6.25 \\
KI \hspace{3mm} & 3.53 \hspace{3mm} & 3.49 \hspace{3mm} & 5.22 \hspace{3mm} & 6.94 \hspace{3mm} & 6.57 \hspace{3mm} & 5.88 \\
RbF \hspace{3mm} & 2.80 \hspace{3mm} & 2.84 \hspace{3mm} & 7.71 \hspace{3mm} & 8.53 \hspace{3mm} & 6.57 \hspace{3mm} & 7.23 \\
RbCl \hspace{3mm} & 3.29 \hspace{3mm} & 3.29 \hspace{3mm} & 6.30 \hspace{3mm} & 7.36 \hspace{3mm} & 6.87 \hspace{3mm} & 6.29 \\
RbBr \hspace{3mm} & 3.44 \hspace{3mm} & 3.43 \hspace{3mm} & 5.75 \hspace{3mm} & 7.06 \hspace{3mm} & 6.82 \hspace{3mm} & 6.03 \\
RbI \hspace{3mm} & 3.68 \hspace{3mm} & 3.64 \hspace{3mm} & 5.14 \hspace{3mm} & 6.65 \hspace{3mm} & 6.87 \hspace{3mm} & 5.68 \\
CsF \hspace{3mm} & 2.98 \hspace{3mm} & 3.05 \hspace{3mm} & 7.64 \hspace{3mm} & 7.94 \hspace{3mm} & 6.56 \hspace{3mm} & 6.73 \\
CsCl \hspace{3mm} & 3.47 \hspace{3mm} & 3.50 \hspace{3mm} & 6.29 \hspace{3mm} & 6.92 \hspace{3mm} & 7.19 \hspace{3mm} & 5.96 \\
CsBr \hspace{3mm} & 3.63 \hspace{3mm} & 3.64 \hspace{3mm} & 5.76 \hspace{3mm} & 6.65 \hspace{3mm} & 7.26 \hspace{3mm} & 5.74 \\
CsI \hspace{3mm} & 3.86 \hspace{3mm} & 3.85 \hspace{3mm} & 5.17 \hspace{3mm} & 6.29 \hspace{3mm} & 7.35 \hspace{3mm} & 5.44 \\
\hline\hline
\end{tabular}
}
\label{table_d}
\end{center}
\end{table}

\subsection{DFT and DFPT}
\label{sec:dft_dfpt}
To show the validity of the hard sphere model for the structure-stability relationship of 2D alkali halides, we performed DFT calculations to calculate the energy difference $\Delta E = E_{\rm hex} - E_{\rm tet}$, where $E_{\rm hex}$ and $E_{\rm tet}$ indicate the total energy (per unit cell) of $AB$ in the hexagonal and tetragonal structures, respectively. The DFT calculations were performed by using Quantum ESPRESSO (QE) \cite{qe}. The computational details are provided in Appendix. The values of $\Delta E$ for the alkali halides $AB$ are plotted in Fig.~\ref{fig_3}. The negative values of $\Delta E$ are observed for the Li-based compounds, which is consistent with our model calculations. 

We next studied the dynamical stability of 2D alkali halides by performing the phonon dispersion calculations within density-functional perturbation theory (DFPT) \cite{dfpt}. The phonon dispersions for 20 $AB$ in the hexagonal and tetragonal structures are shown in Fig.~\ref{fig_4}. For the hexagonal and tetragonal structures, 8 and 16 out of 20 $AB$ were found to be dynamically stable, respectively. Although $E_{\rm hex}$ is larger than $E_{\rm tet}$ for the Li-based compounds, the hexagonal structure was unstable because imaginary phonon frequencies were observed along $\Gamma$-M and $\Gamma$-K lines in the Brillouin zone (BZ). On the other hand, the Li-based compounds in the tetragonal structure is dynamically stable except for the LiI.

Figure \ref{fig_5} shows the electron band structure of 2D alkali halides in the hexagonal and tetragonal structures. Basically, the magnitude of the energy gap decreases when the atoms $A$ and $B$ are changed from Li to Na to K to Rb to Cs and F to Cl to Br to I, respectively, except for a few relationships such as between RbF (CsF) and RbCl (CsCl). A similar trend of the energy gap variation has also been reported in the 2D alkali halides having the planar honeycomb structure, where such a variation was correlated with an increase in the ionic radii \cite{kumar}. 

The conduction band minimum (CBM) of the 2D alkali halides is located at the $\Gamma$ point, while the valence band maximum (VBM) is located at along the $\Gamma$-K line for the hexagonal structure and at the $\Gamma$ or M point or the $\Gamma$-X line for the tetragonal structure. Note that the dispersion curve around the Fermi level is relatively flat, that is, small $k$ dependence of the single-particle energy. As suggested in the 2D II-V compounds such as SrSe \cite{II-V}, ferromagnetism may be induced by the strong Coulomb interaction when the amount of carriers is tuned.  

Table~\ref{table_dft} lists the optimized lattice parameters used in the DFT and DFPT calculations. The values of $a$, $z_A$, and $z_B$ for 2D $AB$ increase with the sum of the ionic radii listed in Table~\ref{table_R}, while the increase in $z_A$ is more moderate than that in $z_B$ when the atom $A$ is fixed. For the cases of NaF, KF, RbCl, and CsBr, the difference between $z_A$ and $z_B$ is small or zero, so that the RbF, CsCl, and CsF show the relationship of $z_A>z_B$. Table~\ref{table_dft} also lists the 2D bulk modulus $B_{\rm 2D}$ (see Appendix for the definition) and the in-plane elastic constants $c_{11}$ and $c_{12}$ \cite{nevalaita,ono2021PRM} by assuming the tetragonal structure. The positive value of $\Delta$ indicates the elastic stability of 2D $AB$. The $B_{\rm 2D}$, $c_{ij}$, and $\Delta$, in turn, decrease with the sum of the ionic radii. The LiF has the largest values of $B_{\rm 2D}$ and $c_{11}$ and the small $c_{12}$, resulting in the largest value of $\Delta$.

\subsection{DFT versus hard sphere model}
\label{sec:dft_hard_sphere}
Table \ref{table_d} lists the values of $d$ and the cohesive energies of $E$ within the DFT and the hard sphere model for the tetragonal structure, where we assumed $d_{\rm DFT}^2=a^2/2 + (z_A-z_B)^2$ because the optimized structure may satisfy $z_A\ne z_B$. The agreement of $d$ between the DFT and model calculations is good, where an averaged relative error is 1.3 \%. However, the Madelung energy of $E_{\rm sp}^{\rm coul}(r)=-\alpha e^2/r$ at $r=d_{\rm sp}$ overestimates the cohesive energy of $E_{\rm DFT}$. To remedy it, we include the core-core repulsive forces due to the Pauli principle into the total energy per ion pair as \cite{AM}
\begin{eqnarray}
 E_{\rm sp}^{\rm tot}(r) = E_{\rm sp}^{\rm coul}(r) + \frac{C}{r^m}.
 \label{eq:etot_sp}
\end{eqnarray} 
By assuming that $E_{\rm sp}^{\rm tot}$ takes a minimum value at $r=d_{\rm sp}$, the parameter $C$ can be expressed as a function of $d_{\rm sp}$, and the total energy is formally expressed as 
\begin{eqnarray}
 E_{\rm sp}^{\rm tot}(d_{\rm sp}) = \frac{m-1}{m}E_{\rm sp}^{\rm coul}(d_{\rm sp}),
\label{eq:Etot}
\end{eqnarray} 
hence the cohesive energy is smaller by a factor of $(m-1)/m$. The value of $m$ is related to the $B_{\rm 2D}=S(\partial^2 E/\partial S^2)_0$, where $S$ is the total area of the surface, and the subscript ``0'' indicates the derivative at the equilibrium. For the tetragonal structure, the $m$ is written as (see Appendix for the derivation)
\begin{eqnarray}
 m = 1+\frac{4 d_{\rm sp}^2 B_{\rm 2D}}{\vert E_{\rm sp}^{\rm coul} (d_{\rm sp}) \vert }.
 \label{eq:m}
\end{eqnarray} 
In the present work, we calculated the $B_{\rm 2D}$ by using DFT, and estimated the value of $m$ using Eq.~(\ref{eq:m}). As listed in Table \ref{table_d}, the corrected total energies given in Eq.~(\ref{eq:Etot}) agree with the DFT results, where an averaged relative error was reduced from 28 \% for $E_{\rm sp}^{\rm coul}$ to 10 \% for $E_{\rm sp}^{\rm tot}$. Note that the magnitude of $m$ estimated in 2D $AB$ is smaller than that estimated in 3D $AB$; for example, $m=$ 5.88, 6.73, 7.24, and 7.06 for LiF, LiCl, LiBr, and LiI in the NaCl-type structure, respectively \cite{AM}. This implies that the core-core repulsive forces are long-ranged in the 2D systems. 

\begin{figure}
\center
\includegraphics[scale=0.45]{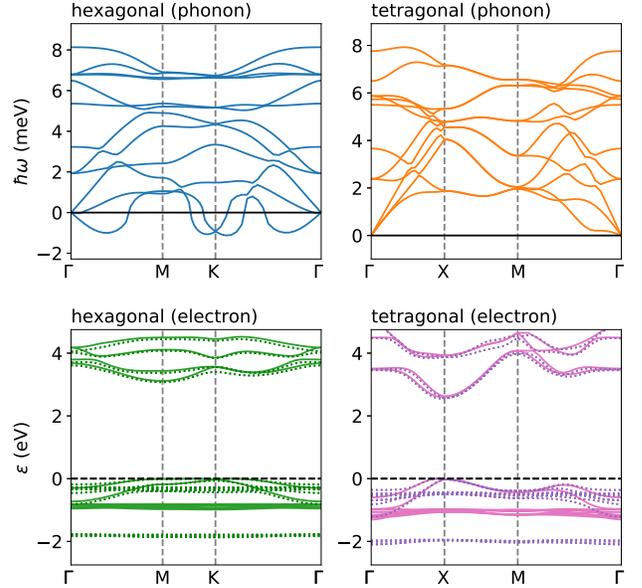}
\caption{The phonon and electron band structures of CsAu in the hexagonal and tetragonal structures. The imaginary phonon frequencies are indicated by negative values. The electron energy is measured from the Fermi energy (horizontal dashed). The electron bands below the Fermi level are split when the SOC is included (dotted). } \label{fig_6} 
\end{figure}

\begin{figure}
\center
\includegraphics[scale=0.4]{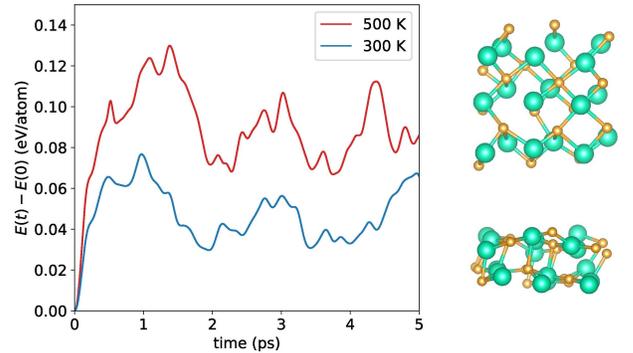}
\caption{Left: The time-evolution of the total energy (per atom) at $T=$ 300 K and 500 K for 2D CsAu in the tetragonal structure. Right: top and side views for the atomic distribution after the first-principles MD simulation of 5 ps for $T=500$ K. } \label{fig_7} 
\end{figure}

\subsection{CsAu}
\label{sec:csau}
As an application of the structural templates in Fig.~\ref{fig_1}, we investigate the stability of 2D CsAu (IA-IB compound). It is known that the Cs and Au atoms form an ionic crystal in the CsCl-type structure. The CsAu shows an optical bandgap of 2.6 eV that was extracted from the photoemission spectroscopy experiments \cite{spicer} and an indirect bandgap of 0.9 eV (from the point X to R in the Brillouin zone) that was predicted by DFT calculations \cite{koenig,CsAu}. The MD simulations have predicted that the liquid CsAu has a frequency gap between the longitudinal and transverse optical branches in the limit of long wavelength, also showing a non-metallic character \cite{bryk}. The 2D CsAu in the buckled honeycomb structure has been predicted to be a wide-gap semiconductor, although the free-standing structure is unstable against the out-of-plane deformations \cite{ono2021PRM}. Such an instability will be avoided by adapting the structural templates for 2D ionic crystals. 

The CsAu compound has the ratio of $R=1.23 < R_{\rm tet}$ by assuming $r=$ 1.37 \AA \ for the Au atom \cite{AM}, so that the tetragonal structure is preferred. In fact, the value of $\Delta E$ was estimated to be 0.15 eV/cell within the DFT. The optimized lattice parameters in units of \AA \ are as follows: $(a, z_A, z_B)=(6.15, 1.79, 1.83)$ and $(5.16, 1.82, 1.84)$ for the hexagonal and tetragonal structures, respectively. For the tetragonal structure, we obtain $d_{\rm DFT}=3.65$ \AA \ that is larger than $d_{\rm sp}=3.06$ \AA. 

Figure \ref{fig_6} shows the phonon dispersions of 2D CsAu in the hexagonal and tetragonal structures, showing that the tetragonal structure is dynamically stable. For the tetragonal phase, we performed first-principles MD simulations by assuming the temperature at $T=300$ K and $500$ K, and confirmed that 2D CsAu is thermodynamically stable as well (see Fig.~\ref{fig_7}). The electron band structure for the 2D CsAu is also shown in Fig.~\ref{fig_6} (bottom panels). The indirect band gap from around K to M point is observed for the hexagonal structure, while the direct band gap at X point is observed for the tetragonal structure. For the hexagonal and tetragonal structures, the magnitude of the energy gap is estimated to be 3.1 eV and 2.6 eV within the generalized gradient approximation \cite{pbe}, respectively, which are higher than that of the 3D CsAu by a factor of more than two. When the spin-orbit coupling (SOC) is included, the lattice parameters of $z_A$ and $z_B$ were almost the same as those without the SOC, while the flat bands around $\varepsilon \simeq -1$ eV split into two bands that are separated by 1.5 eV.

\section{Conclusion}
\label{sec:conclusion}
In conclusion, we investigated the dynamical stability of 2D alkali halides in the hexagonal and tetragonal structures by performing first-principles phonon calculations, and showed that 8 and 16 out of 20 $AB$ were dynamically stable, respectively, indicating that the tetragonal structure can serve as a structural template for the free-standing 2D ionic crystals. The electron energy gaps range from 6.8 eV for LiF to 3.9 eV for RbI and CsI in the tetragonal structure, and a relatively flat band can be observed near the Fermi level. The cohesive energies using a phenomenological model based on the hard spheres with positive or negative charges are in agreement with those using DFT within an average error of 10 \%. The tetragonal CsAu, an ionic compound including only metallic elements, was also predicted to be dynamically and thermodynamically stable and show a band gap of 2.6 eV that is higher than that of the 3D counterparts. The present work is expected to stimulate more investigations for interesting combination of $A$ and $B$ atoms, yielding 2D materials that are partially ionic and partially covalent. Exploring suitable substrates \cite{hennig2014,wang_nacl,yu} for the 2D ionic crystals and a phase transition between 2D structures, as an analog of the B1-B2 transition in 3D systems \cite{florez,toledano}, is left for future work. 

\begin{acknowledgments}
This work was supported by JSPS KAKENHI (Grant No. 21K04628). A part of numerical calculations has been done using the facilities of the Supercomputer Center, the Institute for Solid State Physics, the University of Tokyo and the supercomputer ``Flow'' at Information Technology Center, Nagoya University.
\end{acknowledgments}

\appendix
\begin{table}
\begin{center}
\caption{The lattice constant $a$ (\AA) and the cohesive energy $E_{j}(AB)=\varepsilon_{a}(A)+\varepsilon_{a}(B)-\varepsilon_{j}(AB)$ (eV/(ion pair)) of the 2D $AB$ in the structure $j$s that are planar honeycomb and planar square. }
{
\begin{tabular}{lcccccccccc}\hline\hline
 \hspace{4mm} & honeycomb  \hspace{4mm} & \hspace{4mm} & square \hspace{4mm} &  \\ 
$AB$ \hspace{3mm} & $a$ \hspace{3mm} & $E_{\rm DFT}$ \hspace{3mm} & $a$ \hspace{3mm} & $E_{\rm DFT}$  \\ \hline
LiF \hspace{4mm} & 3.11 \hspace{4mm} & 9.21 \hspace{4mm} & 2.69 \hspace{4mm} & 9.20 \\
LiCl \hspace{4mm} & 3.96 \hspace{4mm} & 7.01 \hspace{4mm} & 3.43 \hspace{4mm} & 6.96 \\
LiBr \hspace{4mm} & 4.24 \hspace{4mm} & 6.28 \hspace{4mm} & 3.68 \hspace{4mm} & 6.22 \\
LiI \hspace{4mm} & 4.65 \hspace{4mm} & 5.50 \hspace{4mm} & 4.04 \hspace{4mm} & 5.41 \\
NaF \hspace{4mm} & 3.73 \hspace{4mm} & 7.92 \hspace{4mm} & 3.16 \hspace{4mm} & 8.04 \\
NaCl \hspace{4mm} & 4.54 \hspace{4mm} & 6.19 \hspace{4mm} & 3.85 \hspace{4mm} & 6.28 \\
NaBr \hspace{4mm} & 4.81 \hspace{4mm} & 5.56 \hspace{4mm} & 4.08 \hspace{4mm} & 5.64 \\
NaI \hspace{4mm} & 5.21 \hspace{4mm} & 4.87 \hspace{4mm} & 4.42 \hspace{4mm} & 4.93 \\
KF \hspace{4mm} & 4.31 \hspace{4mm} & 7.67 \hspace{4mm} & 3.66 \hspace{4mm} & 7.78 \\
KCl \hspace{4mm} & 5.16 \hspace{4mm} & 6.15 \hspace{4mm} & 4.34 \hspace{4mm} & 6.28 \\
KBr \hspace{4mm} & 5.43 \hspace{4mm} & 5.58 \hspace{4mm} & 4.57 \hspace{4mm} & 5.71 \\
KI \hspace{4mm} & 5.84 \hspace{4mm} & 4.94 \hspace{4mm} & 4.90 \hspace{4mm} & 5.07 \\
RbF \hspace{4mm} & 4.55 \hspace{4mm} & 7.47 \hspace{4mm} & 3.87 \hspace{4mm} & 7.56 \\
RbCl \hspace{4mm} & 5.42 \hspace{4mm} & 6.02 \hspace{4mm} & 4.56 \hspace{4mm} & 6.15 \\
RbBr \hspace{4mm} & 5.69 \hspace{4mm} & 5.47 \hspace{4mm} & 4.78 \hspace{4mm} & 5.60 \\
RbI \hspace{4mm} & 6.10 \hspace{4mm} & 4.86 \hspace{4mm} & 5.11 \hspace{4mm} & 4.99 \\
CsF \hspace{4mm} & 4.81 \hspace{4mm} & 7.44 \hspace{4mm} & 4.11 \hspace{4mm} & 7.49 \\
CsCl \hspace{4mm} & 5.71 \hspace{4mm} & 6.03 \hspace{4mm} & 4.82 \hspace{4mm} & 6.14 \\
CsBr \hspace{4mm} & 5.99 \hspace{4mm} & 5.49 \hspace{4mm} & 5.04 \hspace{4mm} & 5.61 \\
CsI \hspace{4mm} & 6.40 \hspace{4mm} & 4.90 \hspace{4mm} & 5.37 \hspace{4mm} & 5.03 \\
\hline\hline
\end{tabular}
}
\label{table_dft2}
\end{center}
\end{table}

\begin{figure*}
\center
\includegraphics[scale=0.5]{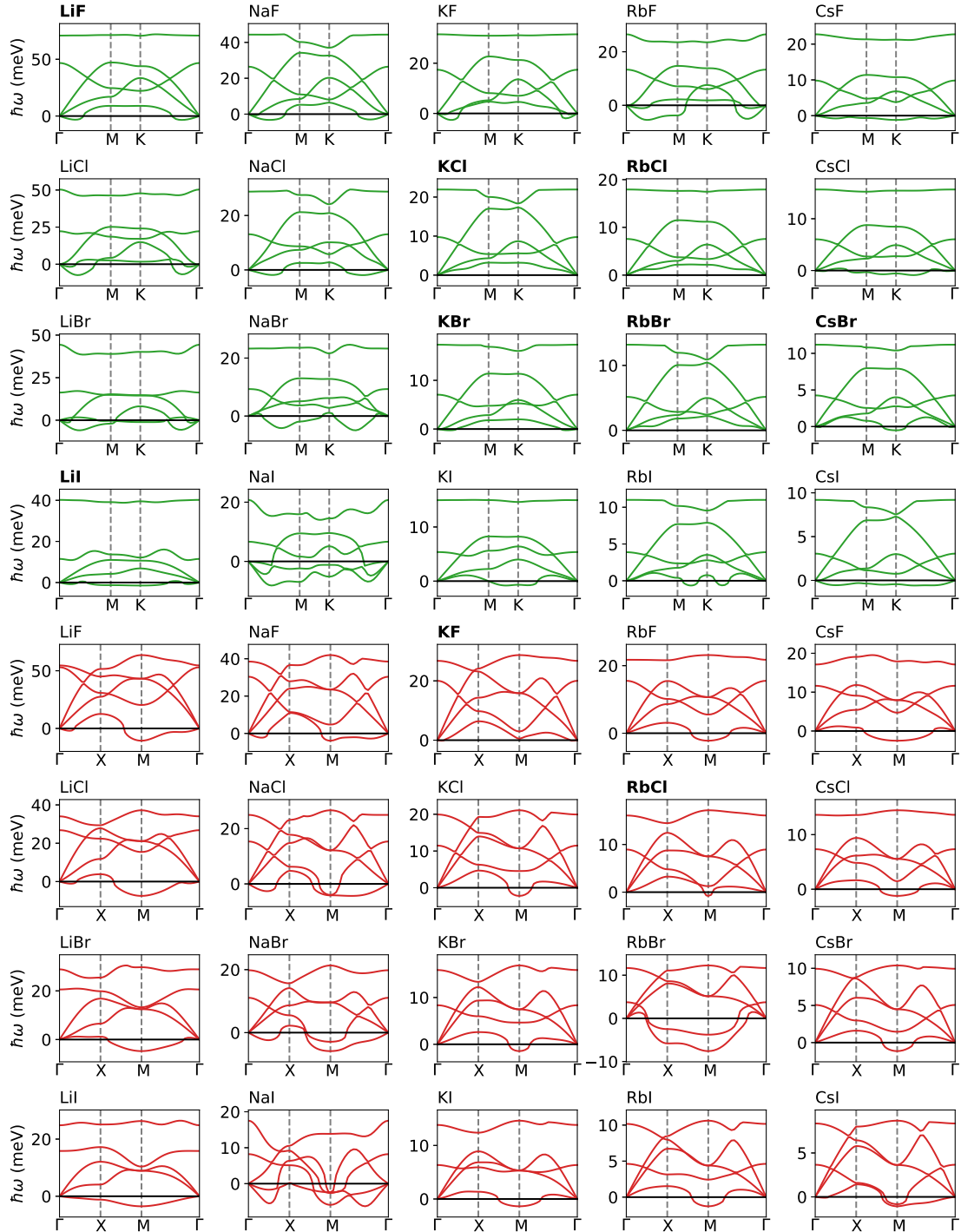}
\caption{The phonon dispersions of 2D alkali halides in the planar honeycomb (green) and planar square (red) structures. The imaginary frequencies are indicated by negative values. The $AB$ is shown in bold when the ratio of the minimum and maximum frequencies is larger than $-0.05$, whereas the honeycomb LiI, CsBr and square RbCl are unstable because imaginary frequencies appear at the K and/or M points. } \label{fig_8} 
\end{figure*}

\section{Computational details}
\label{sec:comp}
We optimized the structural parameters ($a, z_A$, and $z_B$) of 2D ionic crystals by using QE \cite{qe}. The DFT calculations were performed within the generalized gradient approximation of Perdew-Burke-Ernzerhof \cite{pbe}. In the self-consistent field calculations, the ultrasoft pseudopotentials provided in pslibrary.1.0.0 \cite{dalcorso} were used; 60 Ry and 600 Ry were used for the cutoff energies for the wavefunction and the charge density, respectively; and a 20$\times$20$\times$1 $k$ grid was used \cite{MK}. The structures were optimized within an accuracy of $10^{-5}$ Ry for the total energy and $10^{-4}$ a.u. for the total force. The thickness of the vacuum layer was set to be 15 \AA \ to avoid the artificial interactions between the 2D crystals along the $z$ direction. Spin-polarized calculations were performed in the structure optimization, whereas no magnetic moments were observed for the 2D crystals. 

We defined the cohesive energy per ion pair as $E_{\rm DFT}(AB)=\varepsilon_{\rm a}(A) + \varepsilon_{\rm a}(B)-\varepsilon_{\rm tet}(AB)/2$, where $\varepsilon_{\rm a}(X)$ and $\varepsilon_{\rm tet}(AB)$ are the total energy of the atom $X$ and the 2D crystal $AB$ in the tetragonal structure, respectively. To calculate the $\varepsilon_{\rm a}(X)$, we performed atom-in-a-box calculations of a single atom $X$ in a unit cell with a volume of $15\times15\times15$ \AA$^3$ within spin-unpolarized approximation.  

The phonon dispersion calculations were done within DFPT \cite{dfpt}, where a 4$\times$4$\times$1 $q$ grid was used (4 and 6 $q$ points in the BZ for the hexagonal and tetragonal structures, respectively), and the convergence parameter of tr2\_ph was set to be $10^{-14}$. Due to the insulating characteristics of each system, the nonanalytic term accounting for the long-range Coulomb forces is included in the force constant matrix. For the electron band structure calculations, the improved tetrahedron method \cite{tetra_opt} was adapted for creating the charge density. To study the relativistic effect on the electron dispersions of the 2D CsAu, the SOC was included within the scheme described in Refs.~\cite{soc1,soc2}. The first-principles MD simulations were done at 300 K and 500 K for 5 ps, where the velocity scaling method and a time step of 1 fs were used. A 3$\times$3$\times$1 supercell was used for 2D CsAu in the tetragonal structure. Calculations were performed using QE \cite{qe}.

We defined the 2D bulk modulus as 
\begin{eqnarray}
 B_{\rm 2D} = S\left( \frac{\partial^2 E}{\partial S^2}\right)_0,
\end{eqnarray}
where $S$ is the total area of 2D ionic crystals. When there are $N_c$ cells in the crystal, $S=N_c s$ with $s$ being the area per cell. By using the total energy per cell, $e=E/N_c$, the bulk modulus is expressed by $B_{\rm 2D} = s (\partial^2 e/ \partial s^2)_0$. For the tetragonal unit cell with the lattice constant $a$, $s=a^2$ and thus $\delta s \simeq 2a \delta a$. Using the finite difference method, we obtain
\begin{eqnarray}
 B_{\rm 2D} \simeq \frac{e(s+\delta s) + e(s-\delta s) - 2e(s)}{4(\delta a)^2}. 
\end{eqnarray}  
This is evaluated by the total energy calculations within DFT. We next express the $B_{\rm 2D}$ by using the interatomic distance $r$. Assuming $r=a/\sqrt{2}$, $S=2N_c r^2$, and the total number of ions $N=4N_c$, we obtain
\begin{eqnarray}
 B_{\rm 2D} = \frac{1}{4} \left( \frac{\partial^2 E_{\rm sp}^{\rm tot}}{\partial r^{2}}\right)_0,
 \end{eqnarray}
where $E_{\rm sp}^{\rm tot}$ is given by Eq.~(\ref{eq:etot_sp}), and the derivative is evaluated at the equilibrium $r=d_{\rm sp}$. We then obtain the expression of $m$ given by Eq.~(\ref{eq:m}).

The in-plane elastic constants of $c_{11}(=c_{22})$ and $c_{12}$ were calculated by using the formulation described in Refs.~\cite{nevalaita,ono2021PRM}. The elastic stability of the 2D $AB$ was studied by using the determinant of the Hessian matrix of the strain energy, i.e., $\Delta = c_{11}^2-c_{12}^2$. 

We also investigated the dynamical stability of the 2D alkali halides in the planar honeycomb and square structures (i.e., the case of $h=0$) using DFPT \cite{dfpt}, where a 6$\times$6$\times$1 $q$ grid (7 $q$ points) and a 4$\times$4$\times$1 $q$ grid (6 $q$ points) were used, respectively. The optimized lattice parameters, cohesive energies, and phonon dispersions of 20 $AB$ in the planar honeycomb and square structures are provided in Table~\ref{table_dft2} and Fig.~\ref{fig_8}. The 2D alkali halides were unstable except for the planar honeycomb LiF, KCl, KBr, RbCl, RbBr and planar square KF. Such an instabilities may be due to the absence of the ionic bonds along the surface normal, in contrast to the hexagonal and tetragonal structures depicted in Fig.~\ref{fig_1}.



\begin{thebibliography}{99}

\bibitem{heine} P. Miro, M. Audiffred, and T. Heine, An Atlas of Two-Dimensional Materials, Chem. Soc. Rev. {\bf 43}, 6537 (2014).


\bibitem{nevalaita} J. Nevalaita and P. Koskinen, Atlas for the properties of elemental two-dimensional metals, Phys. Rev. B {\bf 97}, 035411 (2018).

\bibitem{ono2020} S. Ono, Dynamical stability of two-dimensional metals in the periodic table, Phys. Rev. B {\bf 102}, 165424 (2020).


\bibitem{tikhomirova} K. A. Tikhomirova, C. Tantardini, E. V. Sukhanova, Z. I. Popov, S. A. Evlashin, M. A. Tarkhov, V. L. Zhdanov, A. A. Dudin, A. R. Oganov, D. G. Kvashnin, and A. G. Kvashnin, Exotic Two-Dimensional Structure: The First Case of Hexagonal NaCl, J. Phys. Chem. Lett. {\bf 11}, 3821 (2020). 

\bibitem{shi} G. Shi, L. Chen, Y. Yang, D. Li, Z. Qian, S. Liang, L. Yan, L. H. Li, M. Wu, and H. Fang, Two-dimensional Na-Cl crystals of unconventional stoichiometries on graphene surface from dilute solution at ambient conditions, Nat. Chem. {\bf 10}, 776 (2018). 

\bibitem{zhao} W. Zhao, Y. Sun, W. Zhu, J. Jiang, X. Zhao, D. Lin, W. Xu, X. Duan, J. S. Francisco, and X. C. Zeng, Two-dimensional monolayer salt nanostructures can spontaneously aggregate rather than dissolve in dilute aqueous solutions, Nat. Commun. {\bf 12}, 5602 (2021). 

\bibitem{cheng} L.-R. Cheng and Z.-Z. Lin, Toward two-dimensional ionic crystals with intrinsic ferromagnetism, Phys. Lett. A {\bf 395}, 127229 (2021). 

\bibitem{sahin} H. \ifmmode \mbox{\c{S}}\else \c{S}\fi{}ahin, S. Cahangirov, M. Topsakal, E. Bekaroglu, E. Akturk, R. T. Senger, and S. Ciraci, Monolayer honeycomb structures of group-IV elements and III-V binary compounds: First-principles calculations, Phys. Rev. B {\bf 80}, 155453 (2009). 

\bibitem{hennig2013} H. L. Zhuang, A. K. Singh, and R. G. Hennig, Computational discovery of single-layer III-V materials, Phys. Rev. B {\bf 87}, 165415 (2013).

\bibitem{hennig2014} A. K. Singh, H. L. Zhuang, and R. G. Hennig, {\it Ab initio} synthesis of single-layer III-V materials, Phys. Rev. B {\bf 89}, 245431 (2014).

\bibitem{cahangirov} S. Demirci, N. Avazl\ifmmode \imath \else \i \fi{}, E. Durgun, and S. Cahangirov, Structural and electronic properties of monolayer group III monochalcogenides, Phys. Rev. B {\bf 95}, 115409 (2017).

\bibitem{II-V} H. Zheng, X.-B. Li, N.-K. Chen, S.-Y. Xie, W. Q. Tian, Y. Chen, H. Xia, S. B. Zhang, and H.-B. Sun, Monolayer II-VI semiconductors: A first-principles prediction, Phys. Rev. B {\bf 92}, 115307 (2015).

\bibitem{IV-VI} C. Kamal, A. Chakrabarti, and M. Ezawa, Direct band gaps in group IV-VI monolayer materials: Binary counterparts of phosphorene, Phys. Rev. B {\bf 93}, 125428 (2016).

\bibitem{IV-V} J.-H. Lin, H. Zhang, X.-L. Cheng, and Y. Miyamoto, Single-layer group IV-V and group V-IV-III-VI semiconductors: Structural stability, electronic structures, optical properties, and photocatalysis, Phys. Rev. B {\bf 96}, 035438 (2017).

\bibitem{ozdamar} B. \"Ozdamar, G. \"Ozbal, M. N. \ifmmode \mbox{\c{C}}\else \c{C}\fi{}\ifmmode \imath \else \i \fi{}nar, K. Sevim, G. Kurt, B. Kaya, and H. Sevin\ifmmode \mbox{\c{c}}\else \c{c}\fi{}li, Structural, vibrational, and electronic properties of single-layer hexagonal crystals of group IV and V elements, Phys. Rev. B {\bf 98}, 045431 (2018).

\bibitem{kumar} P. Kumar, K. Rajput, and D. R. Roy, Structural, vibrational, electronic, elastic and thermoelectric properties of monolayer alkali halide compounds from first principles investigation, Mater. Today Commun. {\bf 29}, 102855 (2021). 

\bibitem{florez} M. Fl\'orez, J. M. Recio, E. Francisco, M. A. Blanco, and A. Mart\'{\i}n Pend\'as, First-principles study of the rocksalt-cesium chloride relative phase stability in alkali halides, Phys. Rev. B {\bf 66}, 144112 (2002). 

\bibitem{toledano} P. Tol\'edano, K. Knorr, L. Ehm, and W. Depmeier, Phenomenological theory of the reconstructive phase transition between the NaCl and CsCl structure types, Phys. Rev. B {\bf 67}, 144106 (2003). 

\bibitem{maradudin} A. A. Maradudin, E. W. Montroll, G. H. Weiss, and I. P. Ipatova, {\it Theory of Lattice Dynamics in the Harmonic Approximation} (Academic Press, New York, 1971).

\bibitem{AM} N. W. Ashcroft, N. D. Mermin, and D. Wei, {\it Solid State Physics} (Cengage Learning, Singapore, 2016).

\bibitem{pettifor} D. Pettifor, {\it Bonding and Structure of Molecules and Solids} (Oxford University Press, New York, 2002).

\bibitem{ashton} M. Ashton, J. Paul, S. B. Sinnott, and R. G. Hennig, Topology-scaling identification of layered solids and stable exfoliated 2D materials, Phys. Rev. Lett. {\bf 118}, 106101 (2017).

\bibitem{balendhran} S. Balendhran, S. Walia, H. Nili, S. Sriram, and M. Bhaskaran, Elemental Analogues of Graphene: Silicene, Germanene, Stanene, and Phosphorene, Small {\bf 11}, 640 (2015).

\bibitem{ono_satomi} S. Ono and H. Satomi, High-throughput computational search for two-dimensional binary compounds: Energetic stability versus synthesizability of three-dimensional counterparts, Phys. Rev. B {\bf 103}, L121403 (2021).

\bibitem{ono2021PRM} S. Ono, Comprehensive search for buckled honeycomb binary compounds based on noble metals (Cu, Ag, and Au), Phys. Rev. Materials {\bf 5}, 104004 (2021).

\bibitem{C2DB2021} M. N. Gjerding, A. Taghizadeh, A. Rasmussen, S. Ali, F. Bertoldo, T. Deilmann, U. P. Holguin, N. R. Kn{\o}sgaard, M. Kruse, A. H. Larsen {\it et al.}, Recent Progress of the Computational 2D Materials Database (C2DB), 2D Materials {\bf 8}, 044002 (2021). 

\bibitem{pymatgen} S. P. Ong, W. D. Richards, A. Jain, G. Hautier, M. Kocher, S. Cholia, D. Gunter, V. L. Chevrier, K. A. Persson, and G. Ceder, Python Materials Genomics (pymatgen): A robust, open-source python library for materials analysis, Comput. Mater. Sci. {\bf 68}, 314 (2013).

\bibitem{qe} P. Giannozzi, O. Andreussi, T. Brumme, O. Bunau, M. B. Nardelli, M. Calandra, R. Car, C. Cavazzoni, D. Ceresoli, M. Cococcioni {\it et al.}, Advanced capabilities for materials modeling with Quantum ESPRESSO, J. Phys.: Condens. Matter {\bf 29}, 465901 (2017).

\bibitem{dfpt} S. Baroni, S. Gironcoli, A. D. Corso, and P. Giannozzi, Phonons and related crystal properties from density-functional perturbation theory, Rev. Mod. Phys. {\bf 73}, 515 (2001). 



\bibitem{spicer} W. E. Spicer, A. H. Sommer, and J. G. White, Studies of the Semiconducting Properties of the Compound CsAu, Phys. Rev. {\bf 115}, 57 (1959). 

\bibitem{koenig} C. Koenig, N. E. Christensen, and J. Kollar, Electronic properties of alkali-metal---gold compounds, Phys. Rev. B {\bf 29}, 6481 (1984).  

\bibitem{CsAu} B. Erdinc, F. Soyalp, and H. Akkus, First-principles investigation of structural, electronic, optical and dynamical properties in CsAu, Cent. Eur. J. Phys. {\bf 9}, 1315 (2011). 

\bibitem{bryk} T. Bryk and I. Klevets, {\it Ab initio} study of collective dynamics in the liquid phase of the equimolar alloy CsAu: Evidence for a nonmetallic state, Phys. Rev. B {\bf 87}, 104201 (2013). 


\bibitem{pbe} J. P. Perdew, K. Burke, and M. Ernzerhof, Generalized Gradient Approximation Made Simple, Phys. Rev. Lett. {\bf 77}, 3865 (1996).



\bibitem{wang_nacl} L. Wang, J. Chen, S. J. Cox, L. Liu, G. C. Sosso, N. Li, P. Gao, A. Michaelides, E. Wang, and X. Bai, Microscopic Kinetics Pathway of Salt Crystallization in Graphene Nanocapillaries, Phys. Rev. Lett. {\bf 126}, 136001 (2021). 

\bibitem{yu} Z. Yu, S. Meng, and M. Liu, Viable substrates for the honeycomb-borophene growth, Phys. Rev. Materials {\bf 5}, 104003 (2021). 


\bibitem{dalcorso} A. Dal Corso, Pseudopotentials periodic table: From H to Pu, Computational Material Science {\bf 95}, 337 (2014).

\bibitem{MK} H. J. Monkhorst and J. D. Pack, Special points for Brillouin-zone integrations, Phys. Rev. B {\bf 13}, 5188 (1976).

\bibitem{tetra_opt} M. Kawamura, Y. Gohda, and S. Tsuneyuki, Improved tetrahedron method for the Brillouin-zone integration applicable to response functions, Phys. Rev. B {\bf 89}, 094515 (2014). 

\bibitem{soc1} A. Dal Corso, Projector augmented-wave method: Application to relativistic spin-density functional theory. Phys. Rev. B {\bf 82}, 075116 (2010).

\bibitem{soc2} A. Dal Corso, Projector augmented wave method with spin-orbit coupling: Applications to simple solids and zincblende-type semiconductors. Phys. Rev. B {\bf 86}, 085135 (2012).
























\end{thebibliography}
\end{document}